\documentclass[amsmath,floatfix,twocolumn,superscriptaddress]{revtex4-1}
\usepackage{subfigure}
\usepackage{siunitx}
\usepackage{amssymb}
\usepackage{amsmath}
\usepackage{graphicx}
\usepackage{array}
\usepackage{dcolumn}
\usepackage{psfrag}
\usepackage{bm}
\usepackage{color}
\usepackage{multirow}
\usepackage{gensymb}
\usepackage{mathptmx}

\begin{document}

\title{Orientational Disorder and Molecular Correlations in Hybrid Organic-Inorganic Perovskites: From Fundamental 
Insights to Technological Applications}

\author{Carlos Escorihuela-Sayalero}
 \affiliation{Group of Characterization of Materials, Departament de F\'{i}sica, Universitat Polit\`{e}cnica de Catalunya,
    Campus Diagonal-Bes\`{o}s, Av. Eduard Maristany 10--14, 08019 Barcelona, Spain}
 \affiliation{Research Center in Multiscale Science and Engineering, Universitat Politècnica de Catalunya,
    Campus Diagonal-Bes\`{o}s, Av. Eduard Maristany 10--14, 08019 Barcelona, Spain}

\author{Ares Sanuy}
 \affiliation{Group of Characterization of Materials, Departament de F\'{i}sica, Universitat Polit\`{e}cnica de Catalunya,
    Campus Diagonal-Bes\`{o}s, Av. Eduard Maristany 10--14, 08019 Barcelona, Spain}
 \affiliation{Research Center in Multiscale Science and Engineering, Universitat Politècnica de Catalunya,
    Campus Diagonal-Bes\`{o}s, Av. Eduard Maristany 10--14, 08019 Barcelona, Spain}

\author{Luis Carlos Pardo}
 \email{luis.carlos.pardo@upc.edu}
 \affiliation{Group of Characterization of Materials, Departament de F\'{i}sica, Universitat Polit\`{e}cnica de Catalunya,
    Campus Diagonal-Bes\`{o}s, Av. Eduard Maristany 10--14, 08019 Barcelona, Spain}
 \affiliation{Research Center in Multiscale Science and Engineering, Universitat Politècnica de Catalunya,
    Campus Diagonal-Bes\`{o}s, Av. Eduard Maristany 10--14, 08019 Barcelona, Spain}

\author{Claudio Cazorla}
 \email{claudio.cazorla@upc.edu}
 \affiliation{Group of Characterization of Materials, Departament de F\'{i}sica, Universitat Polit\`{e}cnica de Catalunya,
    Campus Diagonal-Bes\`{o}s, Av. Eduard Maristany 10--14, 08019 Barcelona, Spain}
 \affiliation{Research Center in Multiscale Science and Engineering, Universitat Politècnica de Catalunya,
    Campus Diagonal-Bes\`{o}s, Av. Eduard Maristany 10--14, 08019 Barcelona, Spain}

\begin{abstract}
{\bf Abstract.} Hybrid organic-inorganic perovskites (HOIP) have emerged in recent years as highly promising semiconducting 
materials for a wide range of optoelectronic and energy applications. Nevertheless, the rotational dynamics 
of the organic components and many-molecule interdependencies, which may strongly impact the functional properties 
of HOIP, are not yet fully understood. In this study, we quantitatively analyze the orientational disorder 
and molecular correlations in the archetypal perovskite CH$_{3}$NH$_{3}$PbI$_{3}$ (MAPI) by performing 
comprehensive molecular dynamics simulations and entropy calculations. We found that, in addition to the usual 
vibrational and orientational contributions, rigid molecular rotations around the C--N axis and correlations 
between neighboring molecules noticeably contribute to the entropy increment associated with the temperature-induced 
order-disorder phase transition, $\Delta S_{t}$. Molecular conformational changes are equally infrequent 
in the low-$T$ ordered and high-$T$ disordered phases and have a null effect on $\Delta S_{t}$. Conversely, the 
couplings between the angular and vibrational degrees of freedom are substantially reinforced in the high-$T$ 
disordered phase and significantly counteract the phase-transition entropy increase resulting from other factors. 
Furthermore, the tendency for neighboring molecules to be orientationally ordered is markedly local, consequently 
inhibiting the formation of extensive polar nanodomains both at low and high temperatures. 
This theoretical investigation not only advances the fundamental knowledge of HOIP but also establishes physically 
insightful connections with contemporary technological applications like photovoltaics, solid-state cooling and 
energy storage.
\\

\end{abstract}

\maketitle

\section{Introduction}
Hybrid organic-inorganic perovskites (HOIP) are solids with the chemical formula ABX$_{3}$, where A and B--X 
represent organic and inorganic ions, respectively. Analogous to oxide perovskites, HOIP exhibit high-temperature
crystalline phases with cubic symmetry. HOIP have emerged as a promising family of optoelectronic and 
energy materials, with significant potential for use in photovoltaic and light-emitting devices \cite{photo1,
photo2,photo3,photo4}, field-effect transistors \cite{transistor1,transistor2,transistor3}, energy storage \cite{estore1,
estore2,estore3}, and solid-state refrigerators \cite{cool1,cool2,cool3,cool4}, among other technologies. 
Methylammonium lead iodide, CH$_{3}$NH$_{3}$PbI$_{3}$ (MAPI), is an archetypal HOIP that has been extensively 
investigated for solar cells and quantum dots applications \cite{solar1,solar2,solar3,qd1,qd2}. 

A characteristic physical trait of HOIP is that, upon increasing temperature, they undergo order-disorder phase 
transitions involving orientational molecular disorder, which on average gives rise to the high-symmetry cubic lattice
\cite{rot1}. Interestingly, the orientational dynamics of the organic cations may profoundly affect the functional 
properties and lattice dynamics of HOIP \cite{review1,rot2,rot3,rot4,rot5}. To cite few examples, the contribution 
of the molecular CH$_{3}$NH$_{3}$$^{+}$ (MA$^{+}$) rotations to the static dielectric response of MAPI has been 
estimated to be as large as $\sim 40$\% \cite{diel1}. The origin of the hysteresis frequently observed in photocurrent-voltage 
measurements of MAPI-based solar devices, which offers promise for memristors and nonvolatile memory applications
\cite{berruet22}, is also thought to be related to the rotational dynamics of the molecular cations \cite{hyste1}. 
Additionally, the contribution of the orientational MA$^{+}$ degrees of freedom to the caloric response of MAPI, as 
driven by external bias like hydrostatic pressure and electric field shifts, has been shown to be substantial 
\cite{liu16,cazorla24}.

Despite these advancements, there remains a fundamental lack of understanding regarding the correlations between 
organic--organic and organic--inorganic ions, and how these correlations may impact the molecular orientational 
dynamics and, in turn, the functional properties of HOIP. For instance, it has been experimentally shown for 
MAPI that, even at temperatures well below the order-disorder phase transition point, the organic cations remain highly 
disordered and mobile \cite{question1,question2}. Conversely, MA$^{+}$ reorientational motion is largely inhibited 
in mixed-halide hybrid perovskites such as MAPbIBr$_{2}$ and MAPbI$_{2}$Br \cite{question3}. The role of molecular 
correlations in the potential formation of ordered molecular domains also remains debatable \cite{diel1,hyste1,zhang23}. 
Similarly, previous studies assessing the existence of caloric effects in HOIP have mostly considered organic cations 
as independent entities, thus neglecting the likely influence of many-molecule interdependencies on the observed 
solid-state cooling figures of merit \cite{cool1,cool2,cool3,cool4,cazorla24}.

In this study, we delve into the analysis of molecular disorder and likely organic--organic and organic--inorganic ionic 
correlations in HOIP by conducting extensive molecular dynamics (MD) simulations for MAPI. Our quantitative 
investigations focus on evaluating the entropy change associated with the temperature-induced order-disorder phase transition
occurring near room temperature (i.e., between the low-$T$ tetragonal $I4/mcm$ and high-$T$ cubic $Pm\overline{3}m$ phases 
\cite{fransson23,phasediag1}), $\Delta S_{t}$, which is a physically interpretable and readily measurable quantity. We 
also address through MD simulations the question about the possible formation of ordered MA cations nanodomains at finite 
temperatures. Additionally, we establish physically insightful connections between our fundamental findings 
and technological applications of current interest like photovoltaics, solid-state cooling and energy storage.

The organization of this article is as follows. First, we present a detailed description of the simulation approach employed 
for the evaluation of $\Delta S_{t}$. Next, we report our computational results along with some discussions, and subsequently 
provide a summary of the main conclusions. The technical details of our classical MD simulations and entropy calculation 
approach can be found in the Methods section.

\begin{figure*}
\includegraphics[width=\linewidth]{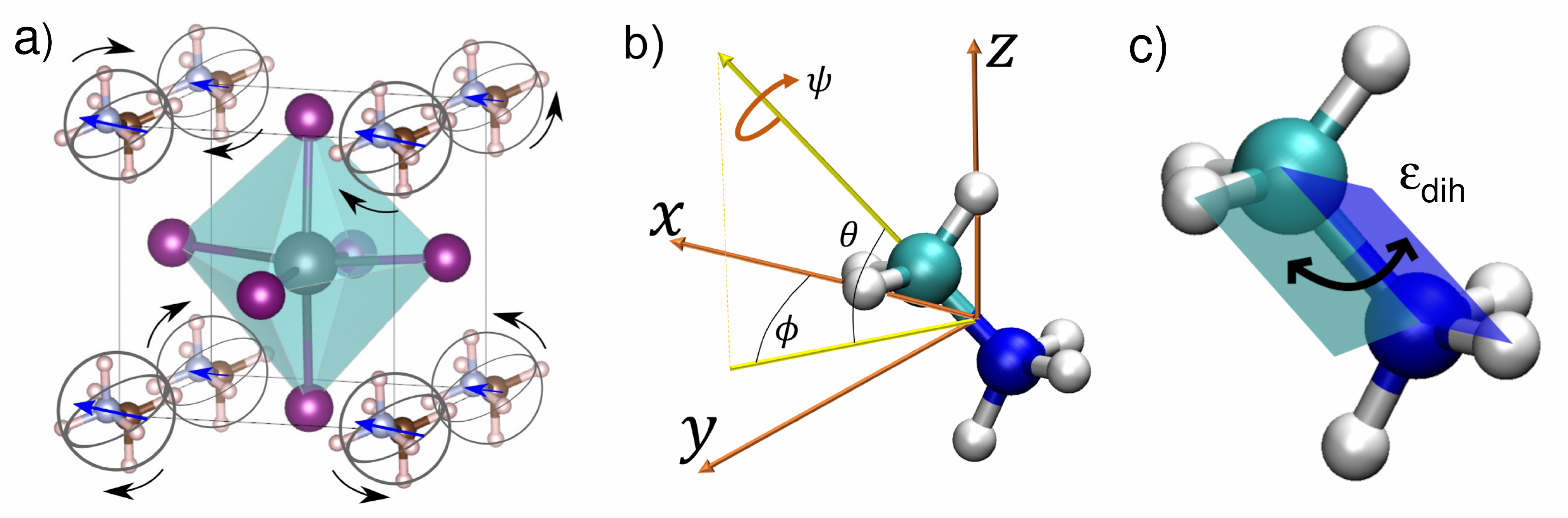}
	\caption{{\bf MAPI unit cell and MA molecule angular coordinates.} (a)~Sketch of MAPI in the high-temperature 
	cubic perovskite phase. MA molecules are orientationally disordered. (b)~Angles definition for describing 
	the orientation of MA molecules. Angles $\theta$ and $\phi$ describe the orientation of the molecular 
	C--N axis. Angle $\psi$ refers to the rotation of the molecule around its C--N axis. (c)~Dihedral angle 
	describing the conformations of a MA molecule. Hydrogen, carbon and nitrogen atoms are represented with white, 
	green and blue spheres, respectively.}
\label{fig1}
\end{figure*}

\section{Computational Approach}
\label{sec:computation}

\subsection{Molecular dynamics simulations}
\label{subsec:md}
Classical molecular dynamics (MD) simulations in the $NpT$ ensemble (i.e., considering fixed number of particles, 
pressure and temperature) have been conducted for bulk MAPI (Fig.\ref{fig1}a) using the atomistic force field developed 
by Mattoni and collaborators \cite{mattoni1,mattoni2}. This classical interatomic potential is based on a hybrid 
formulation of the Lennard--Jones and Buckingham pairwise interaction models, and describes the MA molecules through 
harmonic bonded interactions. Long-range electrostatic interactions are also appropriately accounted for by this atomistic 
force field. The high accuracy of this classical interatomic potential in providing correct MA$^{+}$ orientational 
configurations, as benchmarked by quantum first-principles simulations, has been previously demonstrated 
\cite{cazorla24,kim23} (although it should be noted that it appreciably underestimates the temperature
of the phase transition analyzed in this study, which involves the tetragonal $I4/mcm$ and cubic $Pm\overline{3}m$ 
phases of MAPI \cite{cazorla24,fransson23,phasediag1}). The technical details of our MD--$NpT$ simulations can be 
found in the Methods section.

\subsection{Reference systems and angles definition}
\label{subsec:refsys}
The orientation of each individual MA molecule has been determined in a fixed lab reference system through the 
Euler angles $\theta_{\rm lab}$, $\phi_{\rm lab}$ and $\psi_{\rm lab}$ (Fig.\ref{fig1}b). The angles $\theta_{\rm lab}$ 
and $\phi_{\rm lab}$ determine the orientation of each C--N molecular bond in the fixed reference frame. The angle 
$\psi_{\rm lab}$ describes the rotation of each molecule around its C--N molecular axis. In order to analyse possible 
molecular orientational correlations (see next sections), the relative orientations between MA molecules need to be 
tracked. For this end, a co-mobile reference system is set on an arbitrary molecule and the orientation of the other
molecules are referred to it through the angles $\theta_{\rm rel}$, $\phi_{\rm rel}$ and $\psi_{\rm rel}$. The origin 
of this co-mobile reference system is set at half the distance between the C--N atoms, with the $z$ axis pointing to 
the carbon atom, the $y$ axis being perpendicular to the plane defined by the atoms N--C--H, and the $x$ axis being 
perpendicular to the $y$ and $z$ axes. In addition, a dihedral angle, $\epsilon_{\rm dih}$, formed by the intersecting 
planes containing the H--C--N and C--N--H$^{*}$ atoms (where H belongs to the molecular methyl group and H$^{*}$ to 
the ammonia group, Fig.\ref{fig1}c), is used to monitor possible MA$^{+}$ conformational changes.

\subsection{Entropy calculations}
\label{subsec:entropy}
The entropy of the low-$T$ ordered and high-$T$ disordered phases of MAPI were determined as a function of temperature, 
$S_{\rm total}(T)$, using the relation:
\begin{equation}
	S_{\rm total}(T) = S_{\rm vib}(T) + S_{\rm ang}(T) + S_{\rm vib-ang}(T)~,
\label{eq:stot}
\end{equation}
where $S_{\rm vib}$ is the entropy contribution resulting from the atomic vibrations, $S_{\rm ang}$ from the molecular 
angular degrees of freedom (i.e., orientational and conformational), and $S_{\rm vib-ang}$ accounts for possible couplings 
between the vibrational and molecular angular degrees of freedom. 

Likewise, the molecular angular entropy was assumed to be correctly evaluated with the expression:
\begin{equation}
        S_{\rm ang}(T) = S_{\rm ori}(T) + S_{\rm conf}(T) + S_{\rm ori-ori}(T)~,
\label{eq:sang}
\end{equation}
where $S_{\rm ori}$ is the entropy contribution resulting from the orientational degrees of freedom of a single and 
noninteracting MA molecule, $S_{\rm conf}$ from the conformational changes of a single and noninteracting MA molecule, 
and $S_{\rm ori-ori}$ accounts for possible orientational correlations between different molecules. 

In this study, we primarily have focused on a comparative analysis of the entropy of the low-$T$ ordered and high-$T$ 
disordered phases of MAPI at the corresponding $T$-induced phase transition point (other possible external fields like 
hydrostatic pressure and electric bias are set to zero). Thus, the physical quantity that is quantitatively analyzed 
in detail here corresponds to the phase-transition entropy change, defined as:
\begin{equation}
	\Delta S_{t} = \Delta S_{\rm vib} + \Delta S_{\rm ori} + \Delta S_{\rm conf} +  \Delta S_{\rm ori-ori} 
	             + \Delta S_{\rm vib-ang}~,
\label{eq:deltast}
\end{equation}
where $\Delta S_{\rm x} \equiv S_{\rm x}^{\rm disord} - S_{\rm x}^{\rm ord}$ and all the entropy terms are evaluated at 
the phase-transition temperature $T_{t}$. The main improvement of this $\Delta S_{t}$ definition, compared to that in a 
previous work \cite{cazorla24}, is the inclusion of molecular conformational changes ($\Delta S_{\rm conf}$), molecular 
orientational correlations ($\Delta S_{\rm ori-ori}$), and possible couplings between the vibrational and molecular 
orientational degrees of freedom ($\Delta S_{\rm vib-ang}$).

\subsubsection{Vibrational entropy}
\label{subsubsec:vibrational}
The vibrational density of states $\rho (\omega)$, where $\omega$ represents a lattice vibration frequency, provides information 
on the phonon spectrum of a crystal and allows to estimate key thermodynamic quantities like the vibrational free energy, 
$F_{\rm vib}$, and vibrational entropy, $S_{\rm vib} \equiv -\frac{\partial F_{\rm vib}}{\partial T}$. A possible manner 
of calculating $\rho (\omega)$ from the outputs of MD--$NpT$ simulations consists in estimating the Fourier transform of 
the velocity autocorrelation function (VACF) \cite{cazorla24,sagotra19,cibran23}, defined like:
\begin{equation}
    \text{VACF}(t) = \left\langle \mathbf{v}(0)\cdot \mathbf{v}(t) \right\rangle = \frac{1}{N} \sum_i^{N} \mathbf{v_i}(0)\cdot\mathbf{v_{i}}(t)~,
    \label{eq:vacf}
\end{equation}
where $\mathbf{v_i}(t)$ represents the velocity of the $i$-th particle and $\langle \cdots \rangle$ statistical average 
performed in the $NpT$ ensemble. Subsequently, the vibrational density of states can be estimated like: 
\begin{equation}
    \rho(\omega) = \int_0^{\infty}\text{VACF}(t)~e^{i\omega t}~dt~,
    \label{eq:vdos}
\end{equation}
which fulfills the normalization condition:
\begin{equation}
\int_0^{\infty} \rho(\omega)~d\omega = 3N~,
\label{eq:vdos-norm}
\end{equation}
$3N$ being the number of real and positively defined phonon frequencies of the crystal. 

Upon determination of $\rho$, the vibrational free energy can be straightforwardly estimated with the formula 
\cite{cazorla17}:
\begin{eqnarray}
	F_{\rm vib} (T) & = & k_{B} T \times \nonumber \\  
	& & \int_0^{\infty} \ln{\left[ 2\sinh{\left(\frac{\hbar\omega}{2k_BT}\right)}\right]} \rho(\omega)~d\omega~,
\label{eq:Fvib}
\end{eqnarray}
where $k_{B}$ is the Boltzmann's constant. Consistently, the vibrational entropy adopts the expression:
\begin{eqnarray}
S_{\rm vib} (T) & = & -\int_0^{\infty} k_{B}\ln{\left[2\sinh{\left(\frac{\hbar\omega}{2k_BT}\right)}\right]} \rho(\omega)~d\omega + \nonumber \\ 
& & \int_0^{\infty} \frac{\hbar\omega}{2T} \tanh^{-1}{\left(\frac{\hbar\omega}{2k_BT}\right)} \rho(\omega)~d\omega~.
\label{eq:svib}
\end{eqnarray}

\subsubsection{Molecular orientational entropy}
\label{subsubsec:orientational}
For a continuous random variable $x$ with probability density $f(x)$, its entropy is defined as \cite{informationtheory}:
\begin{equation}
S = - \int_{X} f(x) \log{f(x)}~dx~,
\label{eq:shanon}
\end{equation}
where the integral runs over all possible values of $x$. If $x$ represents a microstate characterizing a particular thermodynamic 
macrostate, then the following Gibbs entropy can be defined for the system of interest \cite{pathria1972}:  
\begin{equation}
S_{G} = -k_{B} \int_{X} f(x) \log{f(x)}~dx~.
\label{gibbs-entropy}
\end{equation}
 
In an orientationally disordered crystal, molecules reorient in a quasi-random manner. By assuming the MA molecules in the 
HOIP crystal to be independent one from the other, one may estimate a probability density function for their orientation, 
$f(\theta_{\rm lab}, \phi_{\rm lab},\psi_{\rm lab})$, from the atomistic trajectories generated during long MD--$NpT$ simulations. 
In this case, the following three-dimensional molecular orientational entropy can be defined, under the implicit assumption
that the length of the MA molecules remains fixed \cite{karplus96,pardo16,pardo15}:
\begin{eqnarray}
	S_{\rm ori}(T) & = &  S_{0}(T) -k_{B} \int f(\theta_{\rm lab}, \phi_{\rm lab}, \psi_{\rm lab}) \log{[f(\theta_{\rm lab}, \phi_{\rm lab}, \psi_{\rm lab})]}  \times \nonumber \\
		    & & d\cos(\theta_{\rm lab})~d\phi_{\rm lab}~d\psi_{\rm lab}~,
\label{eq:sang-cont}
\end{eqnarray}
where $S_{0}(T)$ is a reference entropy term. (For a fluid, the value of this reference entropy term matches that of an ideal 
gas system under the same temperature and density conditions as the system of interest \cite{karplus96,pardo16,pardo15}; 
however, for an orientationally disordered solid, this reference term is not as straightforward to define \cite{cazorla24}.) 

In practice, the calculation of $S_{\rm ori}$ entails the construction of histograms for which the continuous polar variables 
are discretised, $\lbrace \theta_{\rm lab}, \phi_{\rm lab}, \psi_{\rm lab} \rbrace \to \lbrace \theta_{{\rm lab},i}, 
\phi_{{\rm lab},i}, \psi_{{\rm lab},i} \rbrace$. Accordingly, one may define the bin probabilities \cite{informationtheory}:
\begin{eqnarray}
	p_{i}(\theta_{{\rm lab},i},\phi_{{\rm lab},i},\psi_{{\rm lab},i}) \approx f(\theta_{{\rm lab},i},\phi_{{\rm lab},i},
	\psi_{{\rm lab},i}) \times \nonumber  \\
	\Delta\cos(\theta_{\rm lab}) ~\Delta\phi_{\rm lab} ~\Delta\psi_{\rm lab}~,
\label{eq:binprob}       
\end{eqnarray}
where $\Delta\cos(\theta_{\rm lab})~\Delta\phi_{\rm lab}~\Delta\psi_{\rm lab}$ is the volume of a histogram bin (selected 
to be constant in this study). Consistently, one can rewrite the molecular orientational entropy in the discretised form:
\begin{eqnarray}
	\Delta S_{\rm ori}(T) & = & -k_{B} \sum_{i} p_{i} \log{p_{i}} + \nonumber \\ 
	               &   &  k_{B} \log{[\Delta\cos(\theta_{\rm lab})~ \Delta\phi_{\rm lab}~ \Delta\psi_{\rm lab}]}~,
\label{eq:sang-disc}        
\end{eqnarray}
where the value of the reference entropy term in Eq.(\ref{eq:sang-cont}) has been offset.

\subsubsection{Molecular conformational entropy}
\label{subsubsec:conforma}
At finite temperatures, MA molecules may undergo conformational changes that contribute to the total entropy, 
$S_{\rm conf}$. The dihedral angle formed by the intersecting planes containing the H--C--N and C--N--H$^{*}$ 
atoms (where H belongs to the molecular methyl group and H$^{*}$ to the ammonia group, Fig.\ref{fig1}c), $\epsilon_{\rm dih}$, 
may be used to monitor such conformational molecular changes. Analogously to the case of the orientational entropy, 
a probability density function can be estimated for this molecular dihedral angle from the atomistic trajectories 
generated during long MD--$NpT$ simulations, $f(\epsilon_{\rm dih})$. Likewise, a bin probability can be defined, 
$p_{i}(\epsilon_{{\rm dih},i}) \approx f(\epsilon_{{\rm dih},i})~\Delta \epsilon_{\rm dih}$, leading to the 
conformational entropy expression:

\begin{equation}
	\Delta S_{\rm conf}(T) = -k_{B} \sum_{i} p_{i} \log{p_{i}} + k_{B} \log{(\Delta \epsilon_{\rm dih})}~,
\label{eq:conforma}
\end{equation}
where $\Delta \epsilon_{\rm dih}$ is the fixed length of a histogram bin and the value of the corresponding
reference entropy term has been offset.

\subsubsection{Molecular correlation entropy}
\label{subsubsec:correlation}
Equation~(\ref{eq:sang-disc}) would provide the total orientational entropy for the MA molecules if these were 
completely independent one from the other. However, assuming the absence of molecular orientational correlations 
in HOIP might not be a good approximation, so that first these correlations need to be assessed. In this study, 
we account for the effects of molecular correlations in the entropy, $S_{\rm ori-ori}$, by considering up to 
second-order terms in the many-body expansion of the full orientational entropy \cite{karplus96,pardo16,pardo15}. 
To this end, we monitor the relative orientation between pairs of molecules in a co-mobile reference system 
(Sec.\ref{subsec:refsys}), and calculate the molecular correlation entropy difference between the disordered and 
ordered phases like \cite{karplus96,pardo16,pardo15}: 
\begin{equation}
	\Delta S_{\rm ori-ori} = \frac{1}{2} k_{B} \sum_{i=1}^{n_{\rm shell}} n_{i} ~\Delta S^{\rm rel}_{{\rm ori},i}~,
\label{eq:s_oriori}
\end{equation}
where $n_{i}$ is the number of molecules in the $i$-th spherical coordination shell (e.g., $n_{1} = 6$ and 
$n_{2} = 12$), and $n_{\rm shell}$ is the number of considered coordination shells (equal to $5$ in the present 
study). The expression for the individual entropy terms $S^{\rm rel}_{{\rm ori},i}$ is equivalent to that shown in 
Eq.(\ref{eq:sang-disc}), making the angular substitutions $\lbrace \theta_{\rm lab}, \phi_{\rm lab}, 
\psi_{\rm lab} \rbrace \to \lbrace \theta_{\rm rel}, \phi_{\rm rel}, \psi_{\rm rel} \rbrace$.

\begin{figure*}
\includegraphics[width=0.9\linewidth]{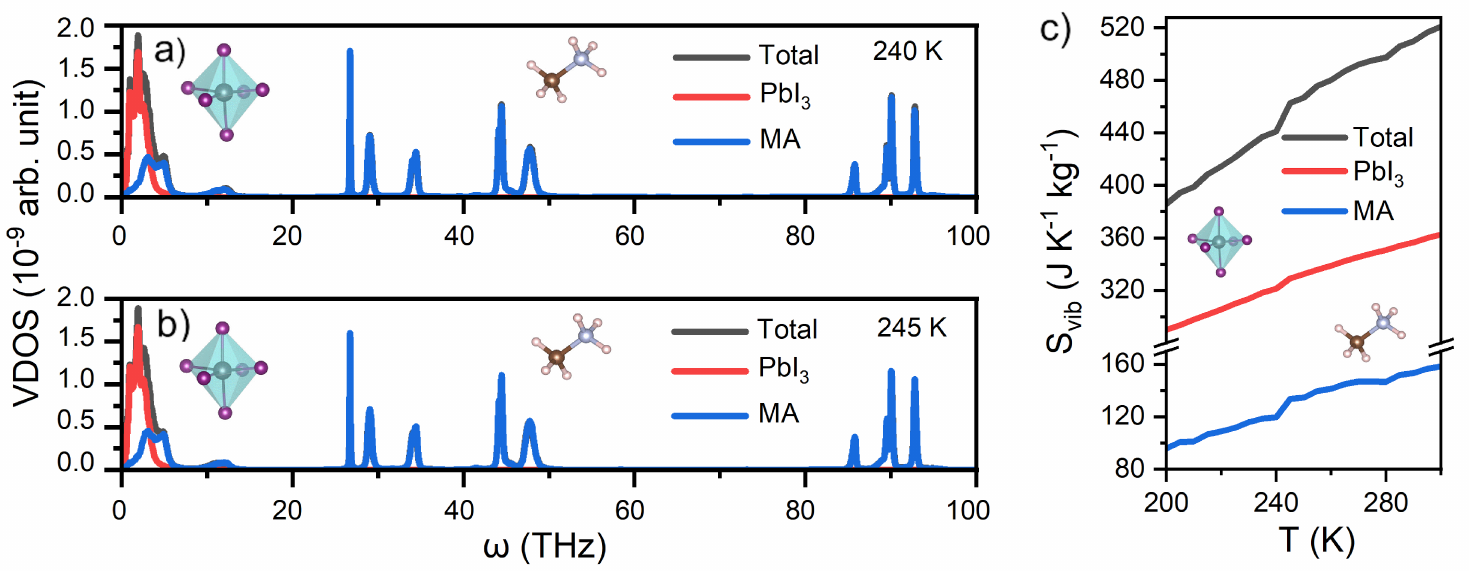}
	\caption{{\bf Vibrational density of states (VDOS) and vibrational entropy ($S_{\rm vib}$) of MAPI calculated 
	at different temperatures.} VDOS results were obtained (a)~for the ordered phase at $T = 240$~K and (b)~for the 
	disordered phase at $T= 245$~K. (c)~Vibrational entropy of MAPI expressed as a function of temperature.}
\label{fig2}
\end{figure*}

It is worth noting that a simulation approach for the calculation of entropy terms, similar to the one introduced in this work,
has been employed in two previous studies on the molecular crystals LiCB$_{11}$H$_{12}$ \cite{plastic1} and C$_{5}$H$_{12}$O$_{2}$
\cite{npg}. In both cases, we observed consistently good agreement for the total phase-transition entropy change when compared
to results obtained using alternative thermodynamic approaches, such as the Clausius-Clapeyron (CC) equation and direct calculation 
of the internal energy difference from MD--$NpT$ simulations (see Sec.~\ref{sec:vib-ang} for further details of these methods).
Thus, despite the limitations in performing additional benchmark calculations to further validate the accuracy of our entropy
estimates for MAPI (e.g., harmonic and thermodynamic integration approaches \cite{thermoint1,thermoint2,brivio15}), due to its
highly anharmonic and orientationally disordered nature, we are confident in the reliability and numerical precision of our
simulation study.

\begin{figure}
\includegraphics[width=0.9\linewidth]{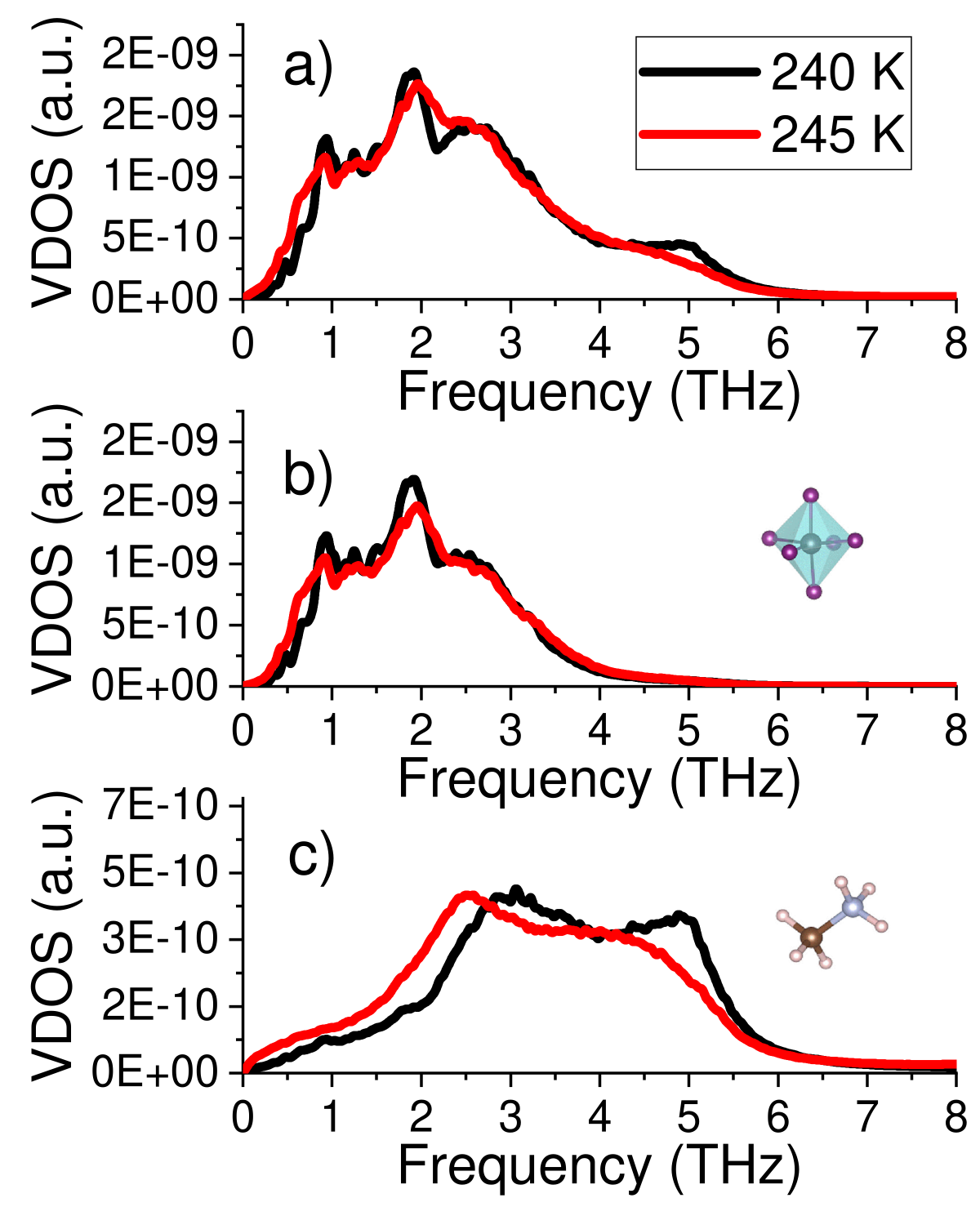}
	\caption{{\bf Vibrational density of states (VDOS) of MAPI in the low-frequency range.} 
	(a)~Total VDOS for the ordered phase at $T = 240$~K and disordered phase at $T= 245$~K. 
	(b)~Inorganic contribution to the total VDOS for the ordered phase at $T = 240$~K and disordered phase at $T= 245$~K. 
	(c)~Organic contribution to the total VDOS for the ordered phase at $T = 240$~K and disordered phase at $T= 245$~K.
	To improve the visualization, a change of scale has been applied on the coordinate axis of this figure.}
\label{fig3b}
\end{figure}

\section{Results and Discussion}
\label{sec:results}
In a previous work \cite{cazorla24}, we already estimated the phase-transition entropy change for MAPI associated 
with the temperature-induced order-disorder phase transition occurring near room temperature (i.e., between the 
low-$T$ tetragonal $I4/mcm$ and high-$T$ cubic $Pm\overline{3}m$ phases \cite{fransson23,phasediag1}), through MD--$NpT$ 
simulations by considering vibrational and molecular orientational degrees of freedom. Nevertheless, in this study, we present 
several critical improvements to our original $\Delta S_{t}$ calculation, which can be generalized to other HOIPs and 
solids exhibiting molecular orientational disorder (e.g., plastic crystals \cite{plastic1,plastic2,plastic3,plastic4,plastic5}). 
These critical computational improvements include considering entropy contributions from molecular conformational 
changes ($\Delta S_{\rm conf}$), molecular orientational correlations ($\Delta S_{\rm ori-ori}$), and possible 
couplings between the vibrational and molecular orientational degrees of freedom ($\Delta S_{\rm vib-ang}$) 
(Sec.\ref{subsec:entropy}). Next, we present and discuss the calculation of $\Delta S_{t}$ term-by-term, comment on 
the potential formation of ordered MA$^{+}$ nanodomains, and make insightful connections with technological 
applications. Our main entropy numerical findings are summarized in Table~\ref{table1}.

\begin{figure}
\includegraphics[width=\linewidth]{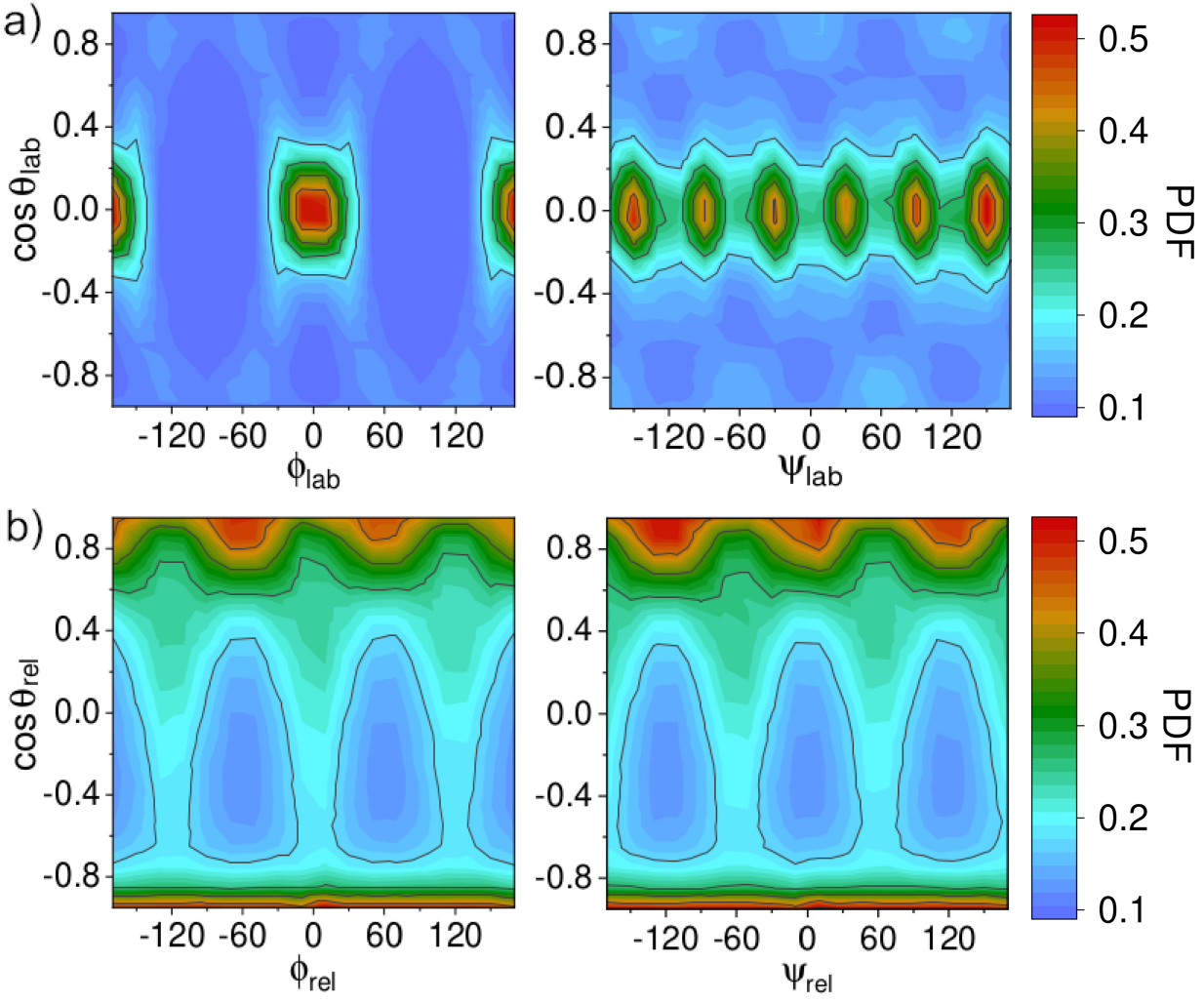}
	\caption{{\bf Bivariate angular probability density function (pdf) for MA molecules in the low-$T$
        ordered phase.} Results were obtained at $T = 240$~K in the (a)~lab-fixed (``lab'') and (b)~molecule-mobile
        (``rel'') reference systems. For the ``rel'' case, the six MA cations within the first coordination shell
	were considered. Red, green and blue colours represent high-probability, medium-probability and low-probability 
	configurations, respectively.}
\label{fig3}
\end{figure}

\begin{figure}
\includegraphics[width=\linewidth]{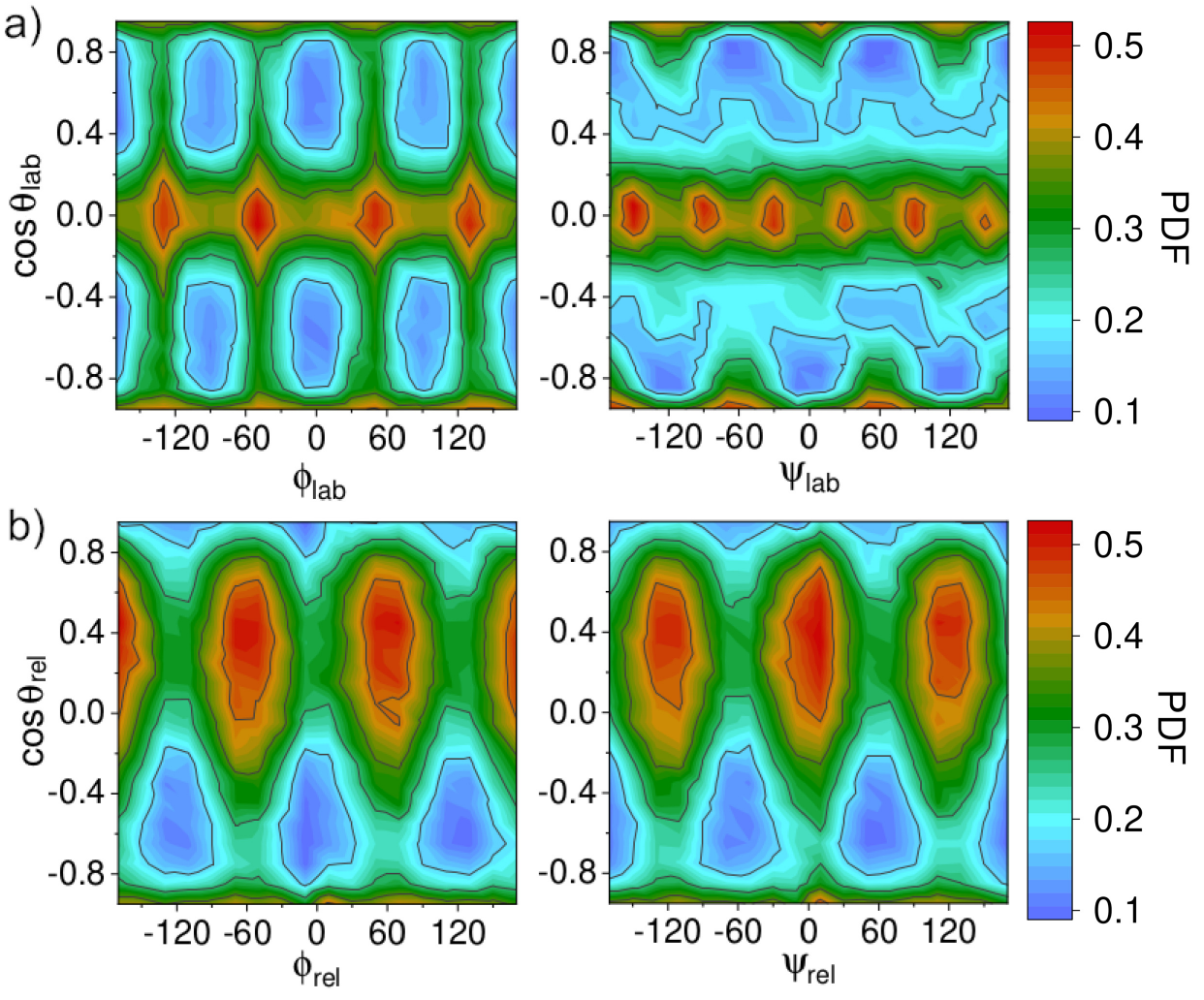}
	\caption{{\bf Bivariate angular probability density function (pdf) for MA molecules in the high-$T$ 
	disordered phase.} Results were obtained at $T = 260$~K in the (a)~lab-fixed (``lab'') and (b)~molecule-mobile 
	(``rel'') reference systems. For the ``rel'' case, the six MA cations within the first coordination shell
        were considered. Red, green and blue colours represent high-probability, medium-probability and low-probability 
	configurations, respectively.}
\label{fig4}
\end{figure}

\subsection{Vibrational entropy}
\label{sec:vibs}
The simulation results presented in this section have been mostly reported in a previous work of ours \cite{cazorla24}. 
These previous numerical findings are reproduced here for coherence and completeness.

Figure~\ref{fig2} shows the vibrational density of states (VDOS) and partial organic and inorganic contributions 
estimated for MAPI at temperatures slightly above and below the simulated order-disorder phase transition point 
(i.e., $245$ and $240$~K). The VDOS contribution corresponding to the inorganic part, namely, the PbI$_{3}$ octahedra, 
is clearly dominant in the low-frequency range $0 \le \nu \le 5$~THz (Figs.~\ref{fig2}a--b). This result follows 
from the fact that the lighter atoms, which typically vibrate at higher frequencies, entirely reside in the organic 
molecules. Consistently, the range of moderate and high frequencies, $\nu \ge 5$~THz, is mostly governed by the 
MA cation vibrations.

Albeit the VDOS enclosed in Figs.~\ref{fig2}a--b may seem quite similar at first glance, there are 
significant differences among them (Fig.\ref{fig3b}a). In particular, the high-$T$ disordered phase accumulates more 
phonon modes in the low-frequency range $0 \le \nu \le 2$~THz than the low-$T$ ordered phase. This effect has a strong 
influence on the vibrational entropy of the system, $S_{\rm vib}$, which is significantly larger for the high-$T$ 
disordered phase. It is worth noting that in previous first-principles computational studies \cite{vib1,vib2}, the 
low-frequency vibrations of the PbI$_{3}$ octahedra (Fig.\ref{fig3b}b) were associated with bending and stretching 
modes, while the low-frequency phonon modes of the MA cations were linked to a combination of libration and 
translation (Fig.\ref{fig3b}c).

Figure~\ref{fig2}c shows the $S_{\rm vib}$ estimated as a function of temperature. A clear surge in vibrational 
entropy is observed at the phase-transition point, amounting to $\Delta S_{\rm vib} = +21.7$~J~K$^{-1}$~kg$^{-1}$.
The positive sign indicates that the vibrational entropy of the high-$T$ disordered phase is larger than that of 
the low-$T$ ordered phase. The primary contribution to this vibrational entropy increase comes from the molecular 
cations, which is equal to $+13.9$~J~K$^{-1}$~kg$^{-1}$ and almost twice as large as that calculated for the 
inorganic part (namely, $+7.8$~J~K$^{-1}$~kg$^{-1}$). Thus, although the low-frequency range in the VDOS is 
dominated by the inorganic anions, the organic MA cations have a larger influence on the vibrational entropy 
change associated with the order-disorder phase transition, $\Delta S_{\rm vib}$.

\begin{figure*}
\includegraphics[width=0.9\linewidth]{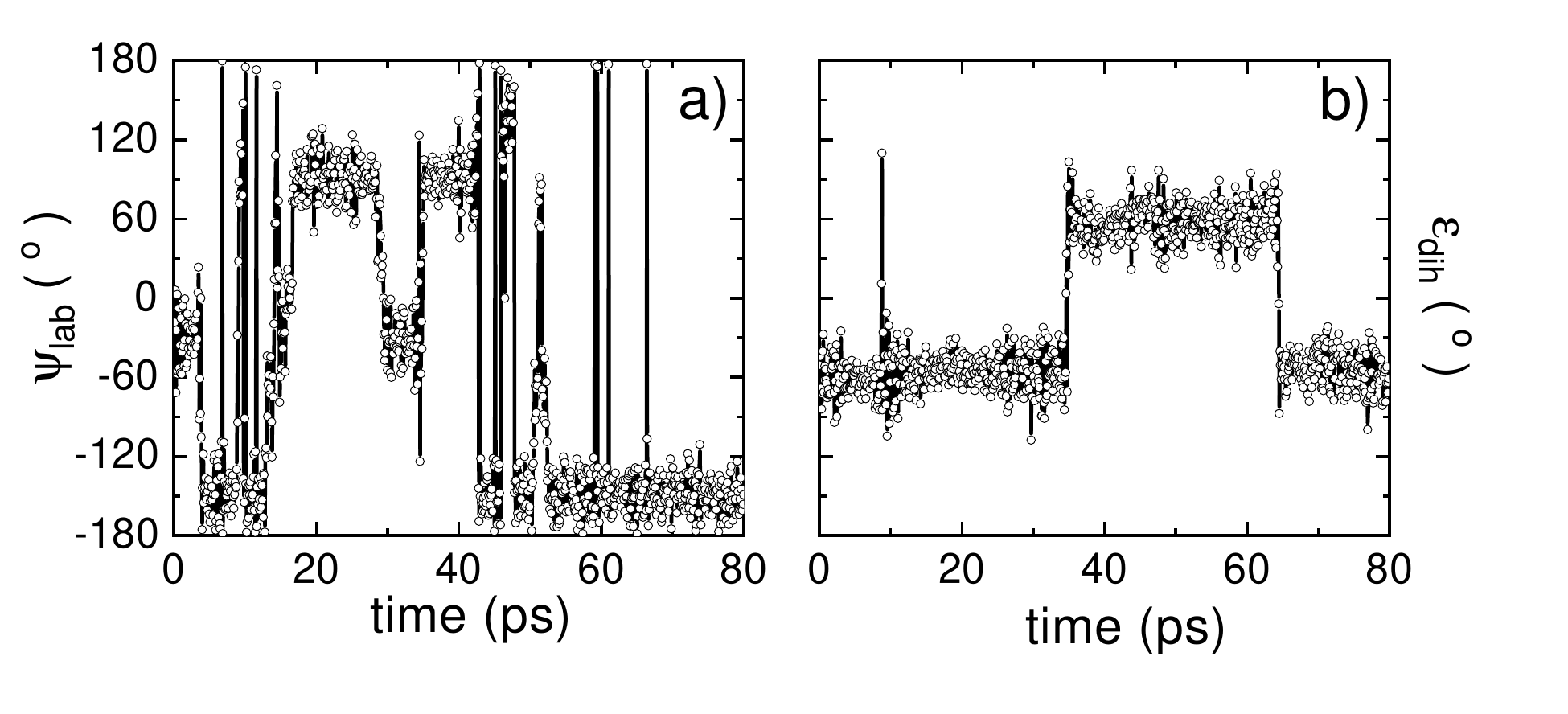} 
        \caption{{\bf Dynamical description of the MA cation orientation around its C--N axis and molecular conformations
        for the low-temperature ordered phase.} Time evolution of (a)~the azimuthal angle $\psi_{\rm lab}$ describing
	the MA$^{+}$ rotation around its C--N axis and (b)~the dihedral angle $\epsilon_{\rm dih}$ describing molecular 
	conformation (Fig.\ref{fig1}).}
\label{fig5}
\end{figure*}

\subsection{Molecular orientational entropy}
\label{sec:oris}
To estimate the orientational entropy associated with the order-disorder phase transition of MAPI, $\Delta S_{\rm ori}$, 
we have extended the computational approach introduced in a previous work of ours \cite{cazorla24} to include the three 
Euler angles that fully determine the orientation of an arbitrary (and rigid) MA cation, namely, $\lbrace \theta, \phi, 
\psi \rbrace$ (Fig.\ref{fig1}b). To calculate $\Delta S_{\rm ori}$, we initially assume the MA cations to be independent 
of each other. Consequently, the three Euler molecular angles should be referred to the stationary lab reference system, 
namely, $\lbrace \theta_{\rm lab}, \phi_{\rm lab}, \psi_{\rm lab} \rbrace$ (Sec.\ref{subsec:refsys}). The complete molecular 
orientation maps thus correspond to three-dimensional probability density functions (pdf) considering those three angles. 

\begin{figure*}
\includegraphics[width=0.9\linewidth]{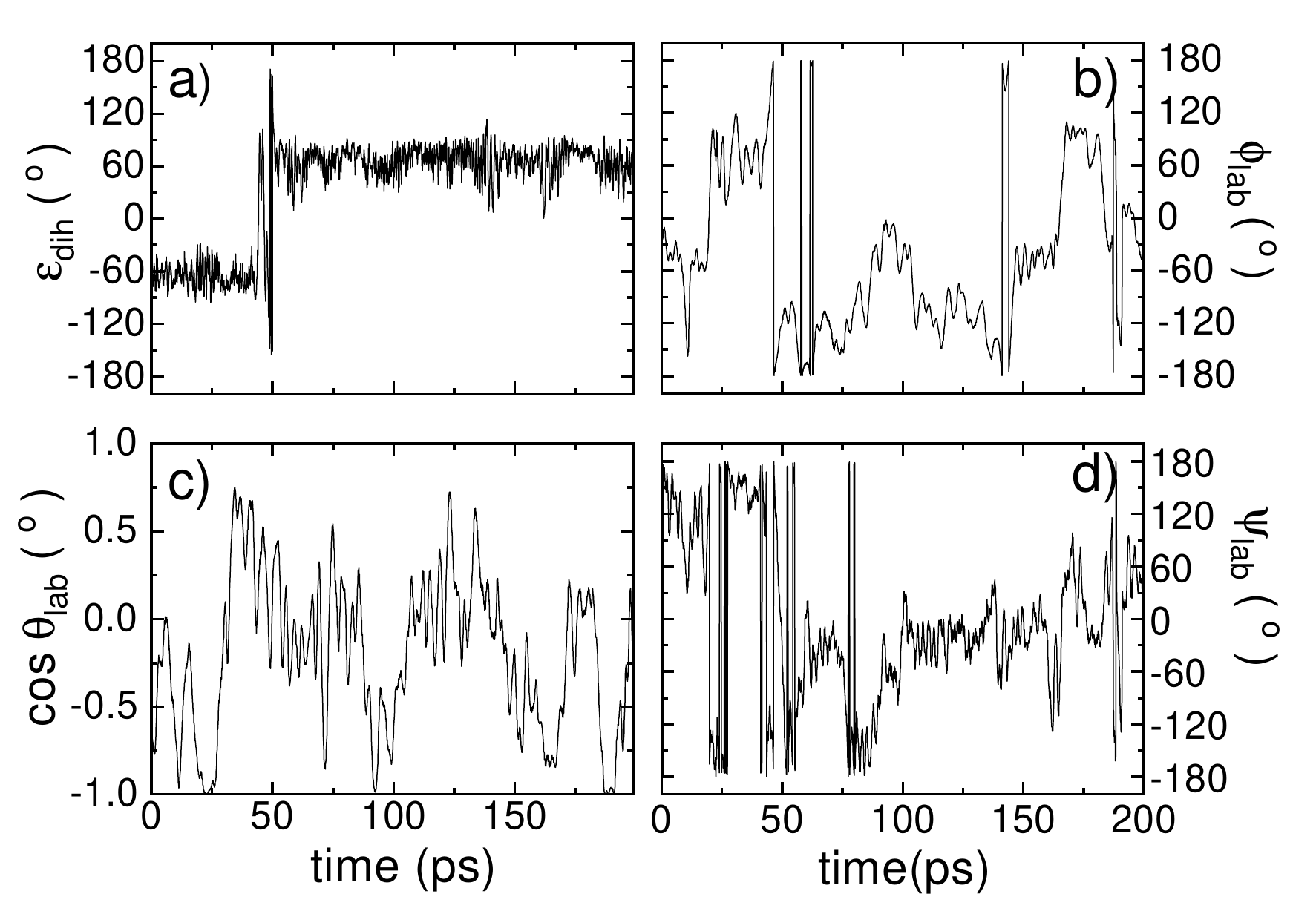}
        \caption{{\bf Dynamical description of the MA cation orientation around its C--N axis and molecular conformations
        for the high-temperature disordered phase.} Time evolution of (a)~the dihedral angle $\epsilon_{\rm dih}$ describing
        molecular conformation, and the rest of MA angles (b)~$\phi_{\rm lab}$, (c)~$\theta_{\rm lab}$ and (d)~$\psi_{\rm lab}$
	describing molecular orientation (Fig.\ref{fig1}).}
\label{fig6}
\end{figure*}

In the previous work \cite{cazorla24}, the $\psi_{\rm lab}$ angle, which accounts for the rotation of the MA molecules 
around their C--N molecular axis (Fig.\ref{fig1}b), was not considered, thus potential entropy contributions resulting 
from this angular degree of freedom were totally neglected. Results for the full MA orientational pdf calculated for 
the low-$T$ ordered and high-$T$ disordered phases of MAPI, as projected onto two different planes, are shown in 
Figs.\ref{fig3}a and \ref{fig4}a, respectively.

As shown in Fig.\ref{fig3}a, in the low-$T$ ordered phase, the MA cations do not reorient since the equilibrium 
molecular orientations, represented by bright spots in the $\cos(\theta_{\rm lab})$--$\phi_{\rm lab}$ pdf, are 
disconnected (i.e., angular transition paths connecting them are absent). This outcome was anticipated and is 
consistent with our previous results reported in \cite{cazorla24}. On the other hand, as expected, in the 
high-$T$ disordered phase, the molecular C--N axis are free to reorient, as indicated by the regions of nonzero 
probability appearing between the equilibrium molecular orientations (Fig.\ref{fig4}a). For this phase, 
and also in agreement with our previous findings \cite{cazorla24}, there a six possible equilibrium molecular 
orientations: two ``apical'' [$\cos(\theta_{\rm lab}) = \pm 1$, $\phi_{\rm lab} = 0$] and four ``equatorial'' 
[$\cos(\theta_{\rm lab}) = 0$, $\phi_{\rm lab} = \pm 60, 120^\circ$].

Figures~\ref{fig3}a and \ref{fig4}a also show the pdf associated with the azimuthal angle $\psi_{\rm lab}$, which 
describes the rigid rotation of the MA cations around their C--N axis (Fig.\ref{fig1}b). Interestingly and surprisingly, 
it is found that for both the low-$T$ ordered and high-$T$ disordered phases of MAPI, there are rotational paths 
connecting high probability regions (i.e., equilibrium molecular orientations). This result implies that, close 
to the transition temperature, the molecules rotate around their C--N axis both in the ordered and disordered states. 
It is worth noting that this type of azimuthal orientational disorder does not break the symmetry of the crystal 
(i.e., the corresponding equilibrium configurations are equivalent among them) and hence cannot be resolved in 
diffraction experiments \cite{azi1,azi2,azi3,azi4}. This class of molecular disorder, present in both 
the low-$T$ ordered and high-$T$ disordered phases of MAPI, is consistent with previous results reported in experimental 
neutron scattering works \cite{rot2,dielphon4}. Nevertheless, in our simulation study, in contrast to experiments, 
we are able to clearly distinguish between rotations of the MA molecule as a whole around the C-N axis, and conformational 
molecular changes involving independent rotations of either the ammonia or methyl group (see next section).

The orientational phase-transition entropy change obtained from the two-dimensional pdf associated with the two angles 
$\lbrace \cos(\theta_{\rm lab}), \phi_{\rm lab} \rbrace$ amounts to $+10.7$~J~K$^{-1}$~kg$^{-1}$ \cite{cazorla24}. 
However, when considering the full orientational pdf associated with the three Euler angles $\lbrace \cos(\theta_{\rm lab}), 
\phi_{\rm lab}, \psi_{\rm lab} \rbrace$, it is found that $\Delta S_{\rm ori} = +13.5$~J~K$^{-1}$~kg$^{-1}$. Therefore,
by including the molecular azimuthal degree of freedom, the orientational entropy change increases by approximately
$25$\%, thus representing a substantial correction. It is noted in passing that the degree of disorder associated with 
the molecular azimuthal angle $\psi_{\rm lab}$ is larger for the high-$T$ phase. 

\begin{figure}
\includegraphics[width=\linewidth]{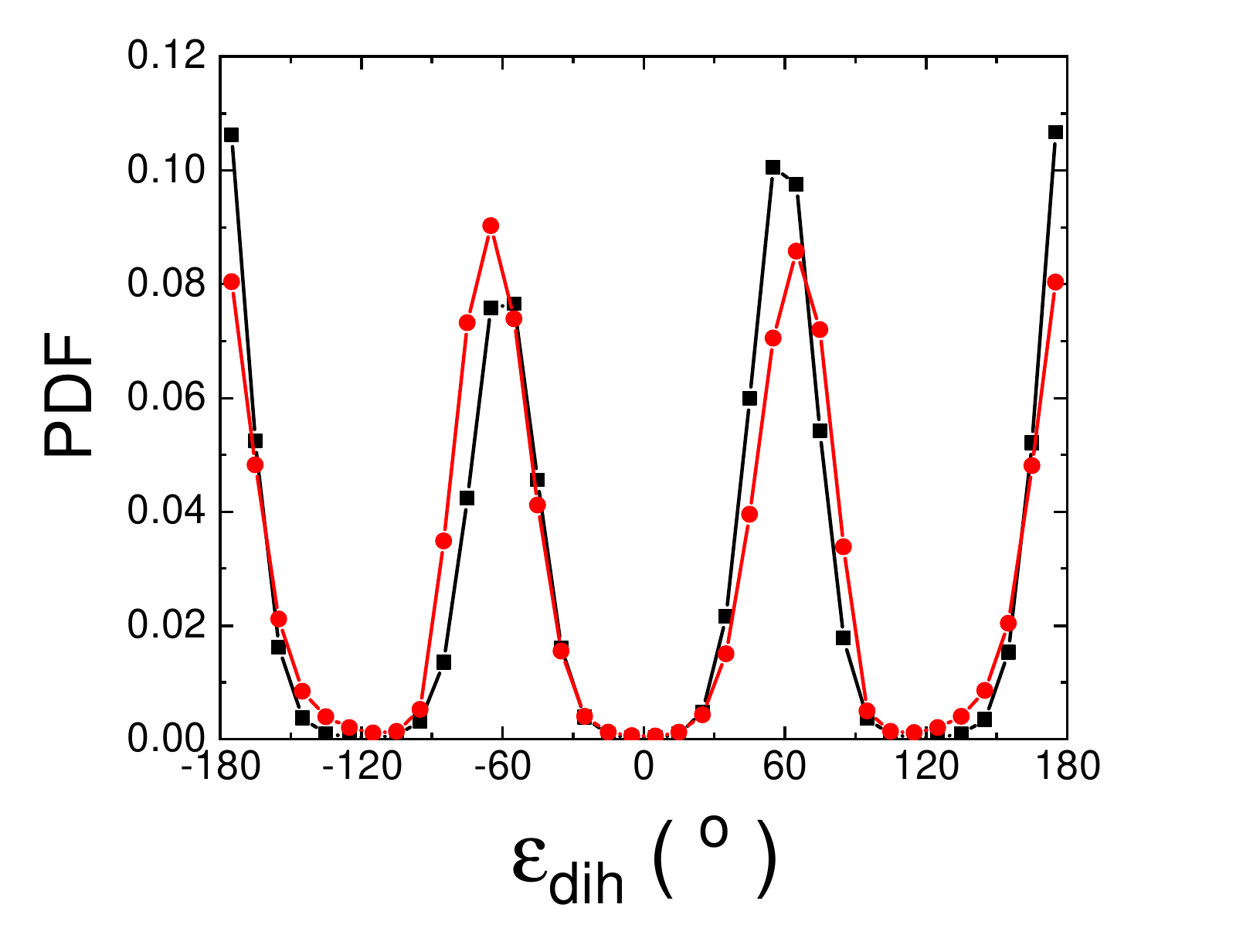}
        \caption{{\bf Angular probability density function (pdf) for the molecular dihedral angle $\epsilon_{\rm dih}$
        describing the conformation of MA molecules.} Results are represented for the low-$T$ ordered (red
        circles) and high-$T$ disordered (black squares) phases. Solid lines are guides to the eye.}
\label{fig7}
\end{figure}

It is instructive to compare the $\Delta S_{\rm ori}$ value obtained from the atomistic MD--$NpT$ simulations with that 
estimated straightforwardly from symmetry and thermodynamic arguments. From the 
$\cos(\theta_{\rm lab})$--$\phi_{\rm lab}$ pdf shown in Figs.\ref{fig3}a and \ref{fig4}a, it is observed that the MA 
cations can adopt two possible orientations in the low-$T$ ordered phase and six in the high-$T$ disordered phase. 
This elementary counting leads to an orientational entropy change of $k_{B} \left( \ln{6} - \ln{2} \right)$, which equals 
$+14.7$~J~K$^{-1}$~kg$^{-1}$ and is approximately $40$\% larger than the $\Delta S_{\rm ori}$ value obtained directly 
from the atomistic simulations. We may thus conclude that such a rough estimation of the orientational entropy change, 
although appears to be a reasonable initial guess, is quantitatively not accurate.

\subsection{Molecular conformational entropy}
\label{sec:confors}
In the previous section, we asserted that in both the low-$T$ ordered and high-$T$ disordered phases of MAPI, the  
MA cations perform rigid rotations around their molecular C--N axis, which do not affect the symmetry of the crystalline 
lattice. However, changes in the azimuthal angle $\psi_{\rm lab}$ could, in principle, also be due to conformational 
changes of the MA cation associated with disconnected rotations of the methyl (CH$_{3}$) and/or ammonia (NH$_{3}$) groups
at the ends of the molecule. To check this possibility, we plot in Figs.\ref{fig5} and \ref{fig6} the time evolution of 
the molecular angles $\lbrace \psi_{\rm lab}, \epsilon_{\rm dih} \rbrace$ for the low-$T$ ordered and high-$T$ disordered 
phases, respectively, where $\epsilon_{\rm dih}$ is the dihedral angle formed by the intersecting planes containing 
the H--C--N and C--N--H$^{*}$ atoms (H belonging to the molecular methyl group and H$^{*}$ to the ammonia group, 
Fig.\ref{fig1}c). For completitude purposes, we also represent the dynamical evolution of the angles $\theta_{\rm lab}$ 
and $\phi_{\rm lab}$ for the high-$T$ disordered phase in Fig.\ref{fig6}.

For the low-$T$ ordered phase (Fig.\ref{fig5}), it is clearly observed that the sequence of $\psi_{\rm lab}$ and 
$\epsilon_{\rm dih}$ changes occurring over time are uncorrelated. For instance, the dihedral angle changes twice 
its value within the time interval of $80$~ps, from $-60^{\circ}$ to $+60^{\circ}$ and viceversa (Fig.\ref{fig5}b), 
while the azimuthal angle changes several tens of times its value, $-180^{\circ} \le \psi_{\rm lab} \le +180^{\circ}$, 
within the same time interval (Fig.\ref{fig5}a). A very similar behaviour is also appreciated for the high-$T$ 
disordered phase (Figs.\ref{fig6}a,d). Consequently, it can be safely concluded that, owing to the absence of time 
correlations between the angles $\psi_{\rm lab}$ and $\epsilon_{\rm dih}$ in both phases, changes in the azimuthal 
angle correspond to rigid molecular rotations around the C--N axis. We note in passing that the dynamics of the 
dihedral angle is extremely slow in both the low-$T$ ordered and high-$T$ disordered phases, which denotes high 
MA$^{+}$ rigidity. 

Figure~\ref{fig7} shows the pdf estimated for the molecular dihedral angle $\epsilon_{\rm dih}$ for both the 
low-$T$ ordered and high-$T$ disordered phases of MAPI, close to the phase-transition temperature. It is observed 
that $\epsilon_{\rm dih}$ essentially adopts the equilibrium values $\pm 60^{\circ}$ and $\pm 180^{\circ}$ in both 
phases. Additionally, the estimated pdf are very similar for the two phases. As a consequence, the phase-transition 
entropy variation associated with molecular conformational changes (Eq.\ref{eq:conforma}) turns out to be practically 
null for MAPI, namely, $\Delta S_{\rm conf} \approx 0$. It can be concluded then that the MA cation presents high 
rigidity in both the low-$T$ ordered and high-$T$ disordered phases, and that below and close to room temperature 
molecular conformational disorder is strongly inhibited in MAPI.

\subsection{Molecular correlation entropy}
\label{sec:corrles}
Figures~\ref{fig3}b and \ref{fig4}b show the pdf estimated for the three Euler angles $\lbrace \theta_{\rm rel}, 
\phi_{\rm rel}, \psi_{\rm rel} \rbrace$ describing the relative orientation between closest MA cations (i.e., 
within the first coordination shell) in a co-mobile molecular reference system (Sec.\ref{subsec:refsys}) for the 
low-$T$ ordered and high-$T$ disordered phases of MAPI, respectively. Using these orientational probability  
maps, and others involving successive coordination shells, it is possible to quantify molecular ordering and 
the phase-transition entropy change associated with the molecule-molecule correlations, as explained in 
Sec.\ref{subsubsec:correlation}.

In the low-$T$ ordered phase (Fig.\ref{fig3}b), the relative orientation between the closest molecules is most 
likely when $\cos(\theta_{\rm rel}) = \pm 1$, which translates into parallel and antiparallel C--N bond arrangements 
or, equivalently, parallel and antiparallel molecular dipoles. Interestingly, it is observed that these maximum 
probability regions differ in shape. The reason for this asymmetry is the following. Two out of the six closest MA 
cations are parallel [$\cos(\theta_{\rm rel}) = +1$], located in the ``tail'' and ``head'' positions of the central 
MA$^{+}$, and four are antiparallel [$\cos(\theta_{\rm rel}) = -1$], positioned in the equatorial plane of the 
central MA$^{+}$. The molecules parallel to the reference MA cation slightly tilt their C--N axis relative to 
the central one to change their $\psi_{\rm rel}$ orientation, causing $\cos(\theta_{\rm rel})$ to slightly depart 
from $+1$. This behaviour, however, is not observed for the antiparallel molecules.   

In the high-$T$ disordered phase (Fig.\ref{fig4}b), strong orientational correlations between closest MA molecules 
are evident, as otherwise, much more uniform and symmetric probability maps would have been obtained. For example, 
molecular parallel arrangements are strongly suppressed at high temperatures, as shown by the very low probability 
estimated for $\cos(\theta_{\rm rel}) = +1$. In this case, the most likely relative MA$^{+}$ orientations correspond 
to three different molecular arrangements resulting in $\cos(\theta_{\rm rel}) \approx 0.4$, which compensate for 
the total dipole moment deriving from the also highly probable antiparallel configuration [$\cos(\theta_{\rm rel}) 
= -1$]. It is worth noting that the number of equilibrium relative orientations is the same for the low-$T$ ordered 
and high-$T$ disordered phases, totaling four. The most significant molecular orientation difference between the two 
phases is a shift of the maximum probability from $\theta_{\rm rel} \approx 0^{\circ}$ at low temperatures to 
$\theta_{\rm rel} \approx 66^{\circ}$ at high temperatures.    

In view of the fact that the reorientational MA$^{+}$ motion in MAPI is correlated, we calculated the entropy difference 
between the low-$T$ ordered and high-$T$ disordered phases resulting from the molecular correlations at the corresponding 
phase-transition temperature, $\Delta S_{\rm ori-ori}$ (Sec.\ref{subsubsec:correlation}). A positive (negative) 
$\Delta S_{\rm ori-ori}$ value would indicate larger (smaller) molecular correlation entropy in the high-$T$ disordered 
phase, and consequently less (more) correlated molecular orientational dynamics in that phase. Adding consecutive 
entropy terms up to the fifth coordination shell (i.e., a maximum radial distance of $14.7$~\AA~ and $56$ molecules), 
we estimated $\Delta S_{\rm ori-ori} = 3.8$~J~K$^{-1}$~kg$^{-1}$. It was found that this molecular correlation entropy 
was already numerically converged to within $0.1$~J~K$^{-1}$~kg$^{-1}$ at the fourth coordination shell, which involves 
a maximum radial distance of $13.2$~\AA~ and $32$ molecules. The calculated $\Delta S_{\rm ori-ori}$ value is positive, 
suggesting more substantial correlations in the molecular orientational dynamics of the low-$T$ phase. Moreover, 
$\Delta S_{\rm ori-ori}$ is comparable in magnitude, for instance, to the phase-transition entropy gain associated with 
the azimuthal angle $\psi_{\rm lab}$ (Sec.\ref{sec:oris}) and hence is not negligible.

\begin{figure*}
\includegraphics[width=0.9\linewidth]{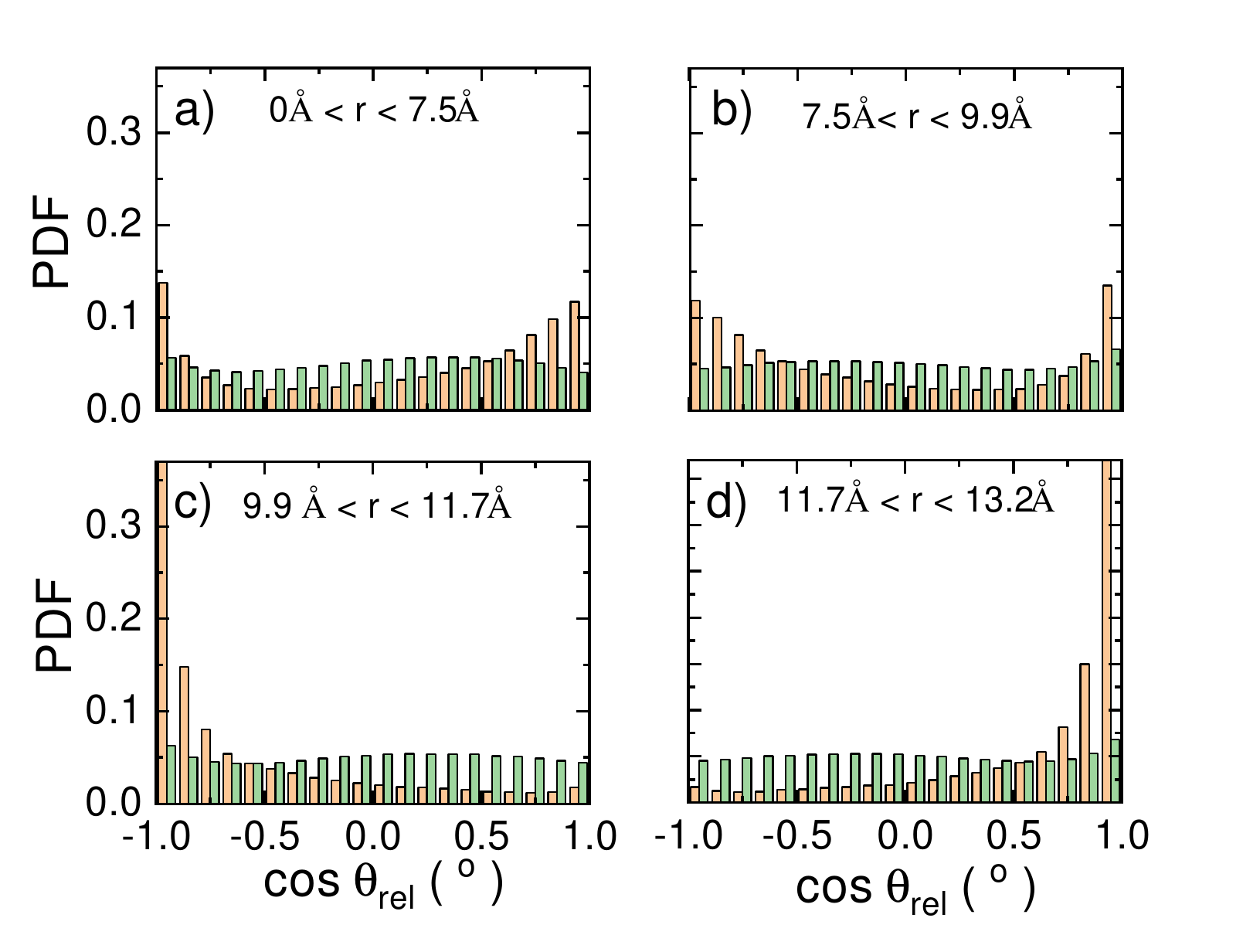}
	\caption{{\bf Probability density function (pdf) for molecular dipole orientation in the co-mobile 
	(``rel'') reference system obtained across successive coordination shells.} Results obtained for the 
	(a)~first, (b)~second, (c)~third and (d)~fourth coordination shells. Orange and green bars represent 
	results obtained for the low-temperature ordered and high-temperature disordered phases, respectively.}
	\label{fig8}
\end{figure*}

\begin{table*}[t]
\begin{tabular}{ccccccc}
\hline
\hline
\qquad  $ $ \quad &  \quad $\Delta S_{\rm vib}$ \qquad & \quad $\Delta S_{\rm ori}$ \qquad & \qquad $\Delta S_{\rm conf}$ \qquad & \qquad $\Delta S_{\rm ori-ori}$ \qquad & \qquad $\Delta S_{\rm vib-ang}$ \qquad  & \qquad $\Delta S_{t}$ \qquad \quad \\
\hline
\qquad (J~K$^{-1}$kg$^{-1}$) &  $+21.7$  & $+13.5$  &  $\approx 0$  &  $+3.8$   & $-12.6$ &  $+26.4$  \\
\qquad (\%)                  &  $+82$    & $+51$    &  $\approx 0$  &  $+15$    & $-48$   &  $+100$   \\
\hline
\hline
\end{tabular}
\caption{\textbf{Contributions to the total entropy change associated with the $T$-induced order-disorder phase transition
        in MAPI, $\Delta S_{t}$.} $\Delta S_{\rm x} \equiv S_{\rm x}^{\rm disord} - S_{\rm x}^{\rm ord}$ and all the entropy
        terms are evaluated at the phase-transition temperature, $T_{t}$. $\Delta S_{\rm vib}$ represents contributions from
        the lattice vibrations, $\Delta S_{\rm ori}$ from the MA orientational degrees of freedom, $\Delta S_{\rm conf}$ from
        the MA conformations, $\Delta S_{\rm ori-ori}$ from the molecule-molecule orientational correlations, and
        $\Delta S_{\rm vib-ang}$ from the couplings between the vibrational and molecular angular degrees of freedom.}
\label{table1}
\end{table*}

\subsection{Molecular dipoles ordering}
\label{sec:molord}
To further elucidate the molecular orientational disorder in MAPI and evaluate the possible formation of ordered 
MA$^{+}$ nanodomains at finite temperatures, we represent in Fig.\ref{fig8} the pdf estimated for the angular 
quantity $\cos(\theta_{\rm rel})$ across successive coordination shells (i.e., from the first up to the fourth).  
Although our primary focus is on the molecular structure of the high-$T$ disordered phase, we have also included 
pdf results for the low-$T$ ordered phase in Fig.\ref{fig8} to facilitate the interpretation and comparison of 
our findings. 

It is clear from Fig.\ref{fig8} that the relative orientation between molecules in the high-$T$ disordered 
phase follows a fairly uniform probability distribution within all the analyzed coordination shells, implying 
that there is no particularly preferred MA$^{+}$ orientation. For instance, the average value of the
angular pdf within the first coordination shell (Fig.\ref{fig8}a) amounts to $\langle \cos(\theta_{\rm rel}) \rangle
= 0.02$, with similar values obtained for the other consecutive coordination shells. Conversely, the pdf 
of the low-$T$ ordered phase exhibits well-defined peaks within all the coordination shells at $\cos(\theta_{\rm rel}) 
= \pm 1$, as expected (Sec.\ref{sec:corrles}), leading to general antipolar ordering.

\begin{figure*}
\includegraphics[width=0.9\linewidth]{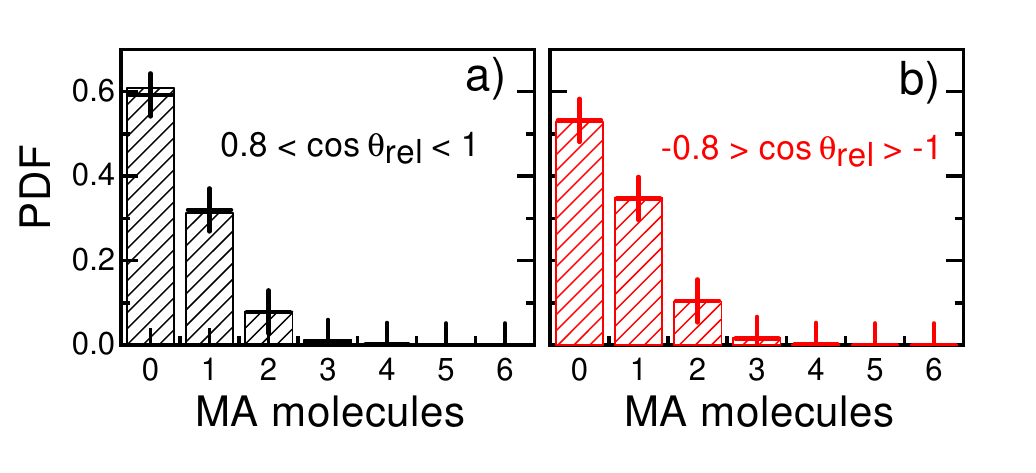}
	\caption{{\bf Probability distribution function (pdf) for the number of molecules in the first 
	coordination shell of a MA cation that are parallel and antiparallel to it.} Results were obtained by averaging 
	over all molecules and simulation time. Pdf corresponding to (a)~parallel and (b)~antiparallel relative molecular 
	dipole orientations (black and red boxes). The crosses in the figures represent analogous results obtained for 
	a random distribution of relative molecular dipole orientations.}
\label{fig9}
\end{figure*}

Interestingly, a very subtle alternation from marginally parallel to marginally antiparallel relative molecular
arrangements is observed for the high-$T$ disordered phase as the radial distance increases. However, this slight
variation of relative MA$^{+}$ orientations developing through successive coordination shells does not induce the
appearance of nanoregions with a well-defined polarization. This outcome is illustrated in Fig.\ref{fig9}, where 
we compare the pdf estimated for the number of parallel and antiparallel MA cations to a chosen one within the 
first coordination shell as directly obtained from our MD--$NpT$ simulations and by considering completely random, 
and thus uncorrelated, relative molecular orientations. No appreciable differences are observed between the two pdf.
It can then be concluded that the appearance of polar nanodomains in MAPI is precluded since the MA molecules
tend to orient homogeneously throughout the crystal. A similar behaviour has been recently reported for a plastic 
crystal \cite{gebbia23}.

\subsection{Vibrational--orientational correlation entropy}
\label{sec:vib-ang}
There is an additional source of potential entropy variation in MAPI related to the correlations between the molecular 
angular and vibrational degrees of freedom, $\Delta S_{\rm vib-ang}$ [Eq.(\ref{eq:deltast})], which mainly results from 
the interdependencies between the organic and inorganic ions. Unlike the rest of terms in Eq.(\ref{eq:deltast}), there is 
neither an analytical expression nor a computational recipe to directly evaluate $\Delta S_{\rm vib-ang}$. Nevertheless, 
this particular entropy variation can be indirectly estimated at the order-disorder phase-transition point using simple 
thermodynamic arguments and MD--$NpT$ simulations, as we explain below. 

At the order-disorder phase-transition temperature, the Gibbs free energy ($G \equiv H - TS$, where $H$ stands for 
enthalpy) of the low-$T$ ordered and high-$T$ disordered phases of MAPI should be equal. Therefore, it follows that 
$\Delta S_{t} = \Delta H_{t} / T_{t}$. Since our MD--$NpT$ simulations are performed at zero pressure, the enthalpy 
difference between the high-$T$ disordered and low-$T$ ordered phases are equal to their internal energy difference
at the phase-transition temperature, namely, $\Delta H_{t} = \Delta U_{t}$. This latter quantity can be directly 
obtained with great precision from the MD--$NpT$ simulations and, consequently, $\Delta S_{t}$ as well. Knowing the
value of all the entropy terms in Eq.(\ref{eq:deltast}) except $\Delta S_{\rm vib-ang}$, it follows that: 
\begin{equation}
        \Delta S_{\rm vib-ang} = \Delta S_{t} - \Delta S_{\rm vib} - \Delta S_{\rm ori} - \Delta S_{\rm conf} - 
	                         \Delta S_{\rm ori-ori}~. 
\label{eq:deltavibang}
\end{equation}

From our MD--$NpT$ atomistic simulations, we obtained $\Delta S_{t} = +26.4$~J~K$^{-1}$~kg$^{-1}$ for the $T$-induced
order-disorder phase transition occurring in MAPI. This value is in fairly good agreement with an equivalent estimation
reported in work \cite{cazorla24}, $\Delta S_{t}^{\rm CC} = +28.4$~J~K$^{-1}$~kg$^{-1}$, which was based on the
Clausius-Clapeyron (CC) method and is therefore more prone to numerical errors. Using Eq.(\ref{eq:deltavibang}) and the
phase-transition entropy change values reported in previous sections, we conclude that $\Delta S_{\rm vib-ang} = 
-12.6$~J~K$^{-1}$~kg$^{-1}$ (Table \ref{table1}). The size of this vibrational--orientational correlation entropy is
roughly $50$\% of the total phase-transition entropy change, which is surprisingly large and comparable in absolute
value to the entropy change contribution $\Delta S_{\rm ori}$ resulting from the individual MA$^{+}$ orientational
degrees of freedom (Sec.\ref{sec:oris}). This source of entropy change in MAPI, and arguably in HOIP in general, has
not been considered in previous studies but, given its large magnitude, should not be neglected.

Interestingly, the sign of the estimated $\Delta S_{\rm vib-ang}$ is negative, unlike the rest of entropy terms in
Eq.(\ref{eq:deltast}). Consequently, the crossed vibrational--orientational contribution decreases the total phase-transition
entropy change. From a physical point of view, the negative sign of $\Delta S_{\rm vib-ang}$ can be understood in terms
of the couplings between the vibrational and molecular angular degrees of freedom, which are substantially reinforced
in the high-$T$ disordered phase (i.e., a smaller entropy is identified here with a higher degree of concertation).
This reasoning is consistent with the large increase in vibrational entropy observed across the order-disorder 
phase transition (Sec.\ref{sec:vibs}). In particular, correlations extending over a large number of atoms, both organic 
and inorganic, may lead to low-frequency phonon modes due to the large effective mass associated with such lattice vibrations 
(recall that for harmonic oscillations, $\omega \propto m^{-1/2}$) \cite{plastic1}. Moreover, the overlap between the 
PbI$_{3}$ and CH$_{3}$NH$_{3}$ lattice excitations also increases in the disordered phase, as compared to that of the 
ordered phase, especially in the limit of very low frequencies (Fig.\ref{fig3b}).

\subsection{Connections to technological applications}

\label{sec:techno}
Table~\ref{table1} reports the value of the different contributions to the total entropy change associated with the 
$T$-induced order-disorder phase transition occurring in MAPI, as well as their relative percentage. The three entropy
variations $\Delta S_{\rm vib}$, $\Delta S_{\rm ori}$ and $\Delta S_{\rm vib-ang}$ are found to be the largest in 
absolute value and similar in size, with the peculiarity that only the sign of the crossed vibrational--orientational 
entropy change is negative. In what follows, we make connections between our atomistic simulation findings and few 
technological applications in which HOIPs are used or have been proposed as promising, namely, solid-state cooling, 
photovoltaics, and energy storage. 

Solid-state cooling is an energy efficient and ecologically friendly technique with potential for solving the 
environmental problems posed by conventional refrigeration technologies relying on compression cycles of greenhouse 
gases \cite{cazorla19,lloveras21,rurali24}. Upon moderate magnetic, electric or mechanical field variations, auspicious caloric 
materials experience large adiabatic temperature variations ($|\Delta T| \sim 1$--$10$~K) due to phase transformations 
entailing large isothermal entropy changes ($|\Delta S| \sim 10$--$100$~J~K$^{-1}$~kg$^{-1}$). HOIP have shown great
promise as caloric materials because their order-disorder phase transition can be driven near room temperature under 
the influence of various external bias like hydrostatic pressure and electric fields \cite{cool1,cool2,cool3,cool4,
zhang23,ren20,liu16,cazorla24}.

However, the entropy changes associated with the order-disorder phase transition in HOIP ($|\Delta S| 
\sim 10$~J~K$^{-1}$~kg$^{-1}$) are about an order of magnitude smaller than those measured in plastic crystals 
($|\Delta S| \sim 100$~J~K$^{-1}$~kg$^{-1}$), a related family of materials composed exclusively of organic molecules. 
Plastic crystals like neopentilglycol (C$_{5}$H$_{12}$O$_{2}$) have recently revolutionized the field of solid-state 
refrigeration due to the huge latent heat associated with their molecular order-disorder phase transition (similar to
that of MAPI), easy synthesis, low toxicity and reduced economic cost \cite{plastic1,plastic2,plastic3,plastic4,plastic5}. 
The physical reason for the difference in phase-transition entropy change between HOIP and plastic crystals may partly 
originate from the $\Delta S_{\rm vib-ang}$ contributions in the former materials, which, as revealed in this study, 
tend to reduce $\Delta S_{t}$. Consequently, a potential rational design strategy to improve the caloric figures of 
merit of HOIP could be to minimize the crossed vibrational--orientational couplings through chemical and/or structural 
engineering \cite{design1}. 

In solar cells, the dielectric function of the light-absorbing materials, $\epsilon$, is of the upmost 
importance as it directly impacts the binding energy of excitons, $E_{\rm bind}$ (i.e., photogenerated electron-hole 
pairs that remain electrostatically bound). In particular, $E_{\rm bind}$ should be minimized to facilitate 
the dissociation of excitons into free charge carriers and prevent electron-hole recombination \cite{zhu22}. According 
to the Wannier-Mott model $E_{\rm bind} \propto 1 / \epsilon^{2}$ \cite{cong23}, meaning larger (smaller) values of 
$\epsilon$ correspond to smaller (larger) exciton binding energies. In HOIP, lattice vibrations are intrinsically 
coupled with cation orientational motion, both of which are considered to influence the material's optoelectronic 
performance \cite{review1,rot2,rot3,rot4,rot5,diel1,hyste1}. Phonons can significantly modulate the band gap, charge 
transport and exciton dynamics \cite{elphon1,elphon2}. Cation orientational dynamics may affect properties such as 
ferroelectricity, ion transport and the dielectric behavior of HOIP \cite{func1,func2,func3}. 

In the specific case of MAPI, $\epsilon$ is strongly influenced by the phonons and molecular cation rotations, which 
both contribute to the lattice component of the dielectric constant in the high-$T$ disordered phase \cite{dielphon1,
dielphon2,dielphon3}, increasing it from $5$ to $33$ \cite{diel1,diel2,diel3}. As a result, the excitonic binding 
energy of MAPI decreases from approximately $16$~meV in the low-$T$ ordered phase to $6$~meV in the high-$T$ disordered 
phase \cite{diel1,diel3}.

In this study, we have theoretically demonstrated that the couplings between phonons and cation orientational dynamics 
in MAPI are substantial and crucial for evaluating the entropy change associated with their $T$-induced order-disorder 
phase transition. Furthermore, by calculating the entropy difference term $\Delta S_{\rm vib-ang}$, we have determined 
that vibrational--orientational couplings are stronger in the high-$T$ disordered phase as compared to those in the 
low-$T$ ordered phase. This enhancement of the molecular cation--ionic anion interactions in the orientationally disordered 
phase may partially explain the observed increase in $\epsilon$ and the corresponding decrease in $E_{\rm bind}$. 
Consequently, a potential rational design strategy to reduce exciton binding energy in light-absorbing HOIPs, thereby 
mitigating detrimental electron-hole recombinations, could involve enhancing their vibrational--orientational couplings, 
either through chemical modifications and/or nanostructuring \cite{design1}. A possible descriptor for quantifying the 
degree of interplay between organic cation orientation and octahedral anion vibration in the high-$T$ disordered phases 
of HOIPs might be the overlap between low-frequency lattice phonons with prevalent molecular and ionic character 
\cite{brivio15}.

Energy storage materials are crucial for powering vehicles, buildings, and portable devices in the push for clean energy. 
Lithium-ion batteries dominate the market, offering high energy density, low self-discharge, negligible memory effect, 
high open-circuit voltage, and durability. Supercapacitors, another key class of energy storage devices, stand out with 
rapid charge/discharge rates, high power density, and long cycle lifetimes. Their capacity surpasses conventional capacitors 
and their faster discharge enables quick electric vehicle charging. Interestingly, halide perovskite materials, originally 
designed for solar cells, have proven effective in energy storage due to their excellent ion diffusion properties (e.g., 
an experimentally measured room-temperature ionic conductivity of $10^{-8}$~S~cm$^{-1}$ for MAPI \cite{zhang15}). While 
ion diffusion was initially considered detrimental to solar cell performance, it is advantageous for lithium batteries 
and supercapacitors, enhancing their efficiency in energy storage applications \cite{estore1,estore2,estore3}.

In MAPI, ionic diffusion, which is sustained by the presence of intrinsic defects, is substantial and increases rapidly with 
rising temperature in the high-$T$ orientationally disordered phase \cite{mattoni2}. In this work, we simulated pristine MAPI 
systems, thus purposely excluding ionic transport. Nevertheless, recent studies in the context of solid-state electrolytes 
have emphasized the significant role of the vibrating non-diffuse crystal matrix in ion migration \cite{sagotra19,cibran23,
ren23,gupta22,muy20,gupta21}. In particular, the interplay between mobile atoms and anharmonic lattice dynamics of the host 
framework may enhance superionicity. 

In this study, the high anharmonicity of the orientationally disordered phase of MAPI has been evidenced by a large 
accumulation of low-frequency phonon lattice modes (Figs.\ref{fig2}--\ref{fig3b}) and a substantial vibrational contribution 
to the total phase transition entropy change, which amounts to approximately $80$\% (Table~\ref{table1}). Potential proxies 
for identifying good ionic conductors among HOIPs may thus include the average phonon frequency, which should be as low as possible, 
along with lattice heat capacity and vibrational entropy, which should be as high as possible \cite{cibran23}. Nonetheless, it 
remains to be determined how the cross vibrational--orientational entropy change term, $\Delta S_{\rm vib-ang}$, is affected 
by the presence of crystalline defects and superionicity, in order to further elucidate the mechanisms of ionic transport 
in HOIPs. These, and others related, interesting questions will be analyzed in detail in future works.

\section{Conclusions}
\label{sec:conclusion}
A comprehensive analysis of the orientational disorder and molecular correlations in the archetypal hybrid organic-inorganic
perovskite (HOIP) MAPI has been presented, relying on atomistic MD--$NpT$ simulations and advanced entropy calculations. 
The main findings of our computational research are the following. Both in the low-$T$ ordered and high-$T$ disordered
phases, there is dynamical orientational disorder associated with rigid MA$^{+}$ rotations around the molecular C--N
axis. This previously overlooked orientational degree of freedom positively contributes to the total phase-transition 
entropy change, $\Delta S_{t}$. The correlations between MA cations are substantial, especially in the low-$T$ ordered 
phase, and have a sizeable augmenting effect on $\Delta S_{t}$. Conformational molecular changes, on the other hand, are 
relatively infrequent both in low-$T$ ordered and high-$T$ disordered phases and do not appreciably contribute to the 
phase-transition entropy change. Interestingly, the couplings between the vibrational and orientational degrees of 
freedom are strengthened in the high-$T$ disordered phase and have a substantial decreasing effect on $\Delta S_{t}$. 
Lastly, molecular correlations in the high-$T$ disordered phase are predominant but markedly local, thus the formation 
of nanoregions with a well-defined polarization is strongly suppressed. These fundamental outcomes may be tentatively 
generalized to other HOIP, and have important ramifications for energy and optoelectronic applications relying on this
family of materials.
\\

\section*{Methods}
\label{sec:methods}
{\bf Molecular dynamics simulations.}~We used the LAMMPS simulation code \cite{lammps} to perform systematic classical 
molecular dynamics simulations in the $NpT$ ensemble for bulk MAPI using the force field developed by Mattoni and 
co-workers \cite{mattoni1,mattoni2}. The average temperature was set using a Nos\'{e}-Hoover thermostat with a mean 
fluctuation of $5$~K. The simulation box contained $3072$ atoms (equivalent to $256$ MAPI unit cells) and periodic 
boundary conditions were applied along the three Cartesian directions. The long-range electrostatic interactions 
were calculated by using a particle-particle particle-mesh solver to compute Ewald sums up to an accuracy of 
$10^{-4}$~kcal~mol$^{-1}$~\AA$^{-1}$ in the atomic forces. The cutoff distance for the evaluation of the potential energy 
was set to $12$~\AA. To determine the phase-transition temperature of MAPI at zero pressure, we conducted comprehensive 
$NpT$-MD simulations in the temperature range $180 \le T \le 340$~K, taken at intervals of $10$~K. In our MD--$NpT$ 
simulations, the temperature was steadily increased up to a targeted value over $1$~ns. Subsequently, the system was 
equilibrated at that selected temperature for $4$~ns. The production runs then lasted for about $1$~ns 
with $\Delta t = 0.5$~fs, from which the velocities of the atoms and other key quantities (e.g., the potential energy and 
volume of the system) were extracted. From the production MD--$NpT$ runs, a total of $1,000$ equispaced configurations
were retrieved to obtain uncorrelated structural data and generate accurate probability density functions (pdf). 
\\

{\bf Molecular angular and entropy analysis.}~The MA$^{+}$ angular degrees of freedom have been retrieved from
the atomic configurations generated during the MD--$NpT$ simulations with the help of the freely available and 
open-source software ANGULA \cite{angula}. ANGULA is designed to automatically and unsupervisedly determine the 
angles defining the orientational structure of molecular disordered crystals from data files containing their 
atomic positions, both in the fixed ``lab'' and co-mobile ``rel'' reference systems. Among its many capabilities, 
ANGULA can generate angular probability density maps and directional radial distribution functions directly from 
sequences of molecular configurations. In this study, the angular molecular entropy terms $S_{\rm ori}$, 
$S_{\rm ori-ori}$ and $S_{\rm conf}$ have been directly computed from the outputs of our MD--$NpT$ simulations 
with ANGULA \cite{angula}. 
\\

\section*{Acknowledgments}
C.C. acknowledges support from the Spanish Ministry of Science, Innovation and Universities under the fellowship
RYC2018-024947-I and grants PID2020-112975GB-I00 and grant TED2021-130265B-C22. The authors also thankfully acknowledge
technical support and computational resources at MareNostrum4 provided by Barcelona Supercomputing Center (FI-2023-1-0002,
FI-2023-2-0004, FI-2023-3-0004, FI-2024-1-0005 and FI-2024-2-0003). This work is part of Maria de Maeztu Units of 
Excellence Programme CEX2023-001300-M funded by MCIN/AEI (10.13039/501100011033).
\\

\newpage

\begin{figure*}
\includegraphics[width=0.8\linewidth]{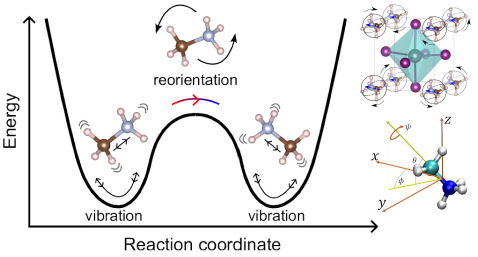}
	\center{{\bf Table of contents (TOC)}}
\label{toc}
\end{figure*}

\end{document}